# Structural and Thermal Stability of $B_4C$/Ru Multilayers with Carbon Barrier Layers

A.V. Bugaev,[a] S.S. Sakhonenkov,[a] A.U. Gaisin,[a] R.A. Shaposhnikov,[b] V.N. Polkovnikov[b] and E.O. Filatova[a]

[a] *Institute of Physics, St-Petersburg State University, Ulyanovskaya Str. 1, Peterhof, St. Petersburg 198504, Russia*

[b] *Institute for Physics of Microstructure, Russian Academy of Sciences, Nizhny Novgorod 603087, Russia*

*Abstract*

The chemical interaction between Mo and Ru layers in multilayer structures depending on the thickness ratio ($\Gamma$) was carried out using X-ray photoelectron spectroscopy (XPS), X-ray diffraction (XRD) and X-ray reflectometry (XRR). The results showed significant interaction of materials inside multilayer structures with the formation of ruthenium borides, with an increase in the $B_4C$ layer thickness (a decrease in the $\Gamma$ parameter) leading to the formation of ruthenium borides of different stoichiometry. The introduction of a carbon barrier layer at the Ru-on-$B_4C$ interface resulted in significant suppression of ruthenium boride formation. The thermal stability of the $B_4C$/Ru system was also studied upon annealing at 400 °C for 1 hour before and after the introduction of the carbon barrier layer. It was shown that the introduction of a carbon barrier layer at the Ru-on-$B_4C$ interface increases the thermal stability of the system, which makes this system more suitable for use in optical systems exposed to long-term radiation. The obtained results are important for the development of highly efficient multilayer mirrors used in EUV lithography and X-ray optics.

*Introduction*

X-ray optics is one of the key technologies in various scientific, engineering and industrial applications such as lithography, high-resolution microscopy, X-ray fluorescence analysis, synchrotrons, free-electron lasers and space astronomy [1–3]. Among these, one of the most critical applications is next-generation lithography. In the microelectronics industry, there is an unceasing global trend of scaling down the manufacturing procedures to increase the operating frequency and decrease the power consumption, and of enhancing the computational capacity of microprocessors [4–6]. Thus, increasing the resolution of the lithography equipment employed in the industry and the associated decrease in the working wavelength of the required light beam is the main technical route agreed upon by industry and science [7].

In fact, lithography at a wavelength of 6.6–6.7 nm (often referred to as Beyond Extreme Ultraviolet (BEUV) lithography) is a promising area of research that can stimulate further development of microchip technology, allowing the creation of more compact, more powerful and more efficient processors. Multilayer mirrors with high reflectivity play a crucial role in the technology of BEUV light management in lithography. This wavelengths range is attractive for the following reasons: firstly, the boron absorption K-edge lies just beyond this spectral range ($\lambda_k$ ~ 6.63 nm). This makes it possible to use boron or its compounds as a weakly absorbing spacer in a multilayer mirror. Secondly, terbium and gadolinium-based light sources [8,9] can provide high system performance. Thirdly, in accordance with the Rayleigh criterion, the transition to this wavelength allows the resolution to be doubled compared to the actively developed lithography with an operating wavelength of 13.5 nm.

It should be noted that other important practical applications of multilayer X-ray mirrors optimized for a working wavelength of 6.7 nm include X-ray fluorescence analysis, which is a non-destructive method for studying various objects [10], such as ores, alloys, water-containing samples, bones and tissues [11,12]. In addition, multilayer X-ray mirrors are actively used as optical elements for synchrotron radiation sources, providing high-intensity monochromatization of the beam [13].

Multilayer structures considered for this wavelength range typically consist of lanthanum or molybdenum paired with $B_4C$ [14–17]. However, theoretical analysis of the reflectance spectra in the wavelengths range from 6 to 9 nm of ideal multilayer X-ray mirrors composed of various materials, highlights the unique potential of multilayer structures based on carbon boride ($B_4C$) and ruthenium (Ru) [18–20]. These systems exhibit outstanding optical properties, making them highly attractive. Ru, as a material with a high atomic number (Z=44), has significant electron density, which is critical to achieving high reflectivity in the soft X-ray range. $B_4C$, in turn, is an ideal partner for Ru due to the optimal combination of low absorption and a suitable refractive index in the working spectral range.

The system's advantages include higher thermal stability and resistance to radiation exposure, which is especially important when working with powerful radiation sources. In [21–23] the effect of annealing on the reflectivity and internal stresses of multilayer Ru/$B_4C$ structures were investigated. It was shown that the reflectivity remains almost unchanged during annealing up to 550 °C, while internal stresses were reduced during thermal annealing.

Despite the high theoretical values of the reflection coefficient, today there is a significant difference between the theoretically predicted and experimentally achieved values, which dictates the need for further thorough studies of the structure in order to identify the reasons for the discrepancy. According to theoretical calculations, the reflectivity of the $B_4C/Ru$ multilayer structure in the wavelength range near to 6.7 nm reaches 70% [24], while the value of experimentally obtained reflectivity does not rise above 30% [25,26].

The main reason for the decrease in reflectivity is the formation of transition layers at the interfaces as a result of the interdiffusion of the main layers with the formation of new compounds [27]. These compounds significantly reduce the contrast of the interfaces and, as a consequence, the efficiency of interference reflection. Another important issue is the increase in interlayer roughness during the growth of the multilayer structure [28]. The accumulation of roughness from layer-to-layer leads to an increase in radiation scattering and a decrease in reflection coherence. Also, the internal stresses found in the $B_4C/Ru$ system, reaching a value of -950 MPa [18], also negatively affect the state of the interfaces between the main layers, as a result, reducing the optical characteristics of these multilayer mirrors.

To overcome the problem of interaction of the layers in multilayer structures, the method of introducing barrier layers at the interfaces is actively used [29,30]. The concept of a barrier layer involves the deposition of an ultra-thin layer (of the order of several angstroms) of material that inhibits interdiffusion between the base layers. In addition, barrier layers can neutralize internal stresses arising in a multilayer structure, which also has a positive effect on the state of the interfaces [18]. With the correct choice of barrier layer material and the interface at which it is introduced, the approach allows maintaining the sharpness of the interfaces, reducing the formation of transition layers and increasing the reflectivity of the structure.

The aim of this work is to study multilayer $B_4C/Ru$ structures obtained by magnetron sputtering. The main attention is paid to the analysis of the formation of interfaces in the system and the influence of the parameter $\Gamma$ (the ratio of the thickness Ru to the period of the structure) on the composition of the transition layer. The prospects for practical application of $B_4C/Ru$ systems are largely related to solving the problem of their long-term stability under operating conditions. Therefore, the thermal stability of the multilayer structure during high-temperature annealing is also studied. The effect of introducing an ultrathin (0.3 nm) C barrier layer at the Ru–$B_4C$ interface is also studied. The results obtained in this work are important for optimizing $B_4C/Ru$ mirrors in EUV lithography and X-ray optics.

*2. Experimental*

All samples were synthesized by direct current magnetron sputtering using the laboratory setup described in the work [31]. Deposition was performed on high-purity super polished Si(100) substrates with an RMS roughness of 0.2 nm in the spatial frequency range of 0.025 – 64 μm$^{-1}$. Before deposition, the pressure of the residual gases in the chamber did not exceed $5 \cdot 10^{-7}$ Torr. During the sputtering process, argon with a purity of 99.998% was used at a working pressure of $2.48 \cdot 10^{-3}$ Torr. Discharge parameters during deposition and the rate of layer growth for the studied multilayer structures are shown in Table 1. The thickness of the layers was controlled by adjusting the speed of the substrate passing over a stationary target.

Multilayer structures are denoted by the formula $[X/Y/Z]_n$, where X, Y, Z are materials of the layers, and n is the number of periods of the structure. The period denotes the total thickness of a group (in base case a pair) of alternating layers. In this notation, the leftmost element (X) corresponds to the layer located closer to the substrate, and the rightmost element (Z) corresponds to the layer located closer to the surface.

Table 1. Discharge parameters during deposition and the rate of layer growth.

| Structure of sample | Material | Voltage, V | Current, mA | Growth rate, nm/s |
|---|---|---|---|---|
| $[B_4C(2.61)/Ru(0.78)]_{50}$ | Ru | 248 | 300 | 0.03 |
| | $B_4C$ | 260 | 1200 | 0.02 |
| $[B_4C(1.75)/Ru(1.44)]_{50}$ | Ru | 238 | 300 | 0.035 |
| | $B_4C$ | 268 | 800 | 0.015 |
| $[B_4C(1.86)/C(0.3)/Ru(1.29)]_{50}$ | Ru | 305 | 300 | 0.1 |
| | $B_4C$ | 299 | 800 | 0.02 |
| | C | 312 | 500 | 0.13 |

The chemical composition of the multilayer structures was studied by X-ray photoelectron spectroscopy (XPS) at the ESCA laboratory module located at the NANOPES station of the Kurchatov Center for Synchrotron Radiation and Nanotechnology. X-ray photoelectron spectroscopy is a non-destructive method for analysis of the elemental composition, as well as the electronic and chemical state of atoms located in the near-surface region of the material under investigation [32,33]. The spectrometer is equipped with hemispherical energy analyzer PHOIBOS HSA 3500 with a diameter of 150 mm, mounted at an angle of 45° to the direction of the X-ray beam. Monochromatized X-ray radiation from an X-ray tube with an aluminum anode (Al $K\alpha$, $hv$ = 1486.6 eV) with an energy resolution better than 0.4 eV is used as the radiation source. The energy calibration of the spectra was performed using the Fermi level, bringing the inflection point to zero. In addition, the spectrometer is equipped with an $Ar^+$ ion gun, which enables to clean sample surface from carbon and oxygen contaminations. When processing the data, the contribution of the inelastic scattering background was taken into account using the universal Tougaard function [34,35]. For a better interpretation, some of the photoelectron spectra were decomposed into components using the Casa XPS software [36].

Data on the crystal structure, layer thickness, and characteristics of the interfaces were obtained by analyzing the results of X-ray diffraction (XRD) and X-ray reflectometry (XRR). The measurements were carried out at a Bruker D8 Discover laboratory diffractometer of the "Research Centre for X-ray Diffraction Studies" of Saint Petersburg State University. The experimental setup includes an X-ray tube with a copper anode and a high-precision horizontal goniometer, which provides scanning in $\theta - 2\theta$ geometry with a sample positioning accuracy of 50 μm and an angular resolution of 0.001°. Theoretical calculations and fitting of experimental reflection curves were performed using the IMD [37] and Multifitting software [38].

### 3. Results and discussion

#### 3.1 Reference structures

As a result of chemical interactions between the main layers in the multilayer $B_4C$/Ru structures, the formation of ruthenium (carbo-)borides can be expected. To date, no reports have been found in the literature describing the formation of ruthenium carbo-borides $RuB_xC_y$ in $B_4C$/Ru systems, while the formation of ruthenium borides has been observed in previous studies [27,28]. Table 2 summarizes the formation energies of several known ruthenium borides and carbides, as well as $B_4C$ obtained from the Materials Project database [39] and from the literature [40]. The data clearly show that Ru-C compounds are thermodynamically unstable under ambient conditions. Among the borides, $RuB_2$ and $Ru_2B_3$ have the most negative formation energies indicating a higher driving force for their formation.

To further assess the likelihood of interfacial product formation, we employed an interface reaction analysis approach similar to that described in [41] and [42], where the thermodynamic driving force for reaction between two solids is evaluated from bulk phase data using convex-hull phase diagrams. In this formalism, all possible reactions between $B_4C$ and Ru were enumerated within the compositional space {B, C, Ru}, and the most exergonic product set was identified. Despite the high intrinsic stability of $B_4C$, the calculations indicate a strong thermodynamic tendency for it to react with Ru to form $Ru_2B_3$ and elemental carbon. The reaction energy is negative (-0.352 eV/atom) relative to the stable phases of Ru and $B_4C$, which implies that the formation of $Ru_2B_3$ at the interface is energetically favorable. Therefore, in the studied systems it is reasonable to expect the occurrence of ruthenium boride phases at the $B_4C$/Ru interfaces.

**Table 2.** Formation energies for ruthenium borides.

| Borides | RuB | $RuB_2$ | $Ru_2B_3$ | $Ru_7B_3$ | RuC | $Ru_2C$ | $B_4C$ |
|---|---|---|---|---|---|---|---|
| Formation energy (eV/atom) | 0.801 | -0.308 | -0.352 | -0.142 | 0.649 | 0.950 | -0.647 |

To study the processes occurring in multilayer $B_4C$/Ru structures, it is necessary to consider the XPS spectra of reference structures. Ru and $B_4C$ films were used for this purpose.

Figure 1a shows the decomposition of Ru $3d_{5/2}$ photoelectron spectrum obtained from the reference 10 nm pure Ru film after surface cleaning with $Ar^+$ ions. Here and below, when the spectra are decomposed into components, the measured spectrum is shown by black dots, and the red line shows the curve representing the sum of the components of the decomposition. Since the Ru $3d_{3/2}$ peak strongly overlaps with the C 1s peak of an adventitious carbon and is considerably separated from the Ru $3d_{5/2}$ peak by about 4.2 eV, only the Ru $3d_{5/2}$ line will be analyzed here and below. The Ru $3d_{5/2}$ photoelectron peak measured for the reference Ru has energy position 279.9 eV and the full width at half maximum (FWHM) 0.4 eV, which correlates well with the literature data [43,44].

Figure 1b shows the B 1s photoelectron spectrum measured for the reference $B_4C$ film. Boron carbide has a complex structure made up of C-B-C chains and $B_{12}$ or $B_{11}C$ icosahedra [45,46]. As a result, the spectrum includes both B-B and B-C bonds. The spectrum decomposition shows that the main peak is a combination of two components: one at 188.4 eV (FWHM = 1.8 eV) corresponding to B-B bonds, and another at 189.7 eV (FWHM = 1.9 eV) corresponding to B-C bonds [47,48]. A small peak from boron oxide is also visible at 191.2 eV.

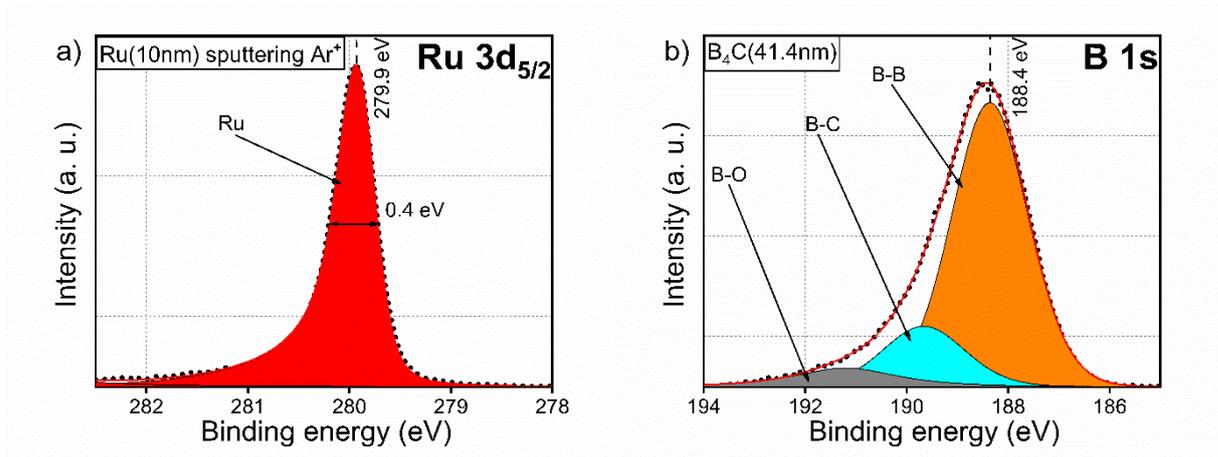

Figure 1. Decomposed Ru $3d_{5/2}$ and B 1s photoelectron spectra obtained from reference Ru(10nm) (a) and $B_4C$(41.4 nm) (b) films.

*3.2 $B_4C$/Ru multilayer structures with different $\Gamma$*

Particular attention should be paid to the parameter $\Gamma$, which determines the ratio of the thickness of the strongly absorbing material (ruthenium) to the thickness of the structure period d. By decreasing the thickness of the absorbing layer and increasing the thickness of the less absorbing one within a fixed value of the parameter, it is possible to reduce the overall absorption losses while maintaining high reflectivity of the system. However, it should not be forgotten that too large decrease in the thickness of the absorbing layer will lead to deterioration of the interference. The balance is achieved at some intermediate values of $\Gamma$ [49,50]. It is also worth remembering that the mutual diffusion between the layers can also be different at different ratios of the layer thicknesses in systems with the same period value. Changing the parameter $\Gamma$ can also affect the transverse and columnar roughness in a multilayer system [49,51]. Thus, studying structures with different $\Gamma$ parameters is important to determine the optimal $\Gamma$ value that minimizes interface imperfections and also reduces absorption. In this regard, two multilayer systems consisting of 50 periods with a nominal period thickness of about 3.3 nm and $\Gamma$ parameters equal to 0.23 and 0.45, respectively, were analyzed.

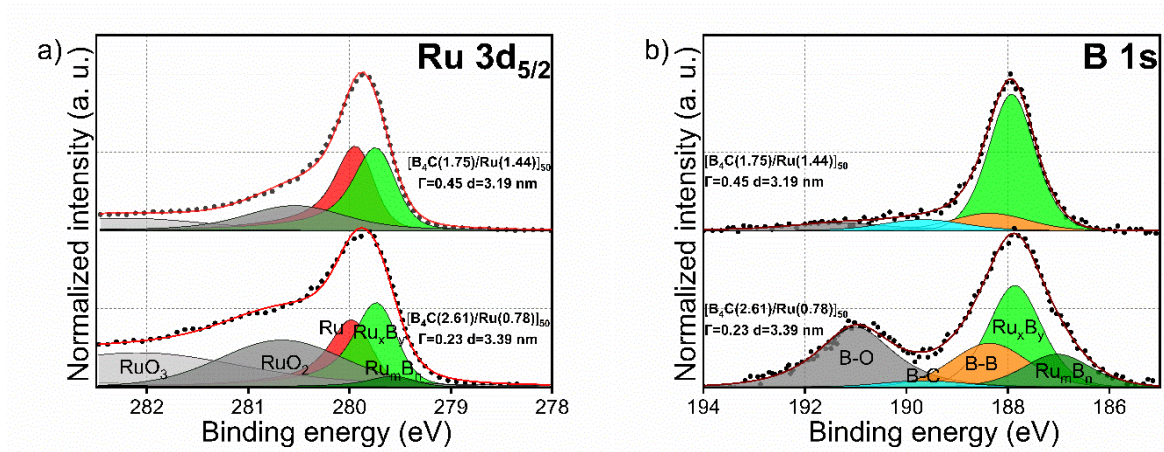

Figure 2. Decomposed Ru 3d$_{5/2}$ (a) and B 1s (b) photoelectron spectra obtained from [B$_4$C/Ru]$_{50}$ multilayer systems with $\Gamma$ = 0.23, 0.45.

The decomposition of the Ru 3d$_{5/2}$ and B 1s photoelectron spectra obtained for the multilayer structures [B$_4$C/Ru] with $\Gamma$ = 0.23, 0.45 is presented in Figure 2. The line shape, energy position and FWHM of the peaks obtained for the reference Ru and B$_4$C were rigidly fixed and used in the decomposition of the spectra of the multilayer structures. Here and below, all photoelectron spectra were normalized to peak intensity for the comparison convenience.

Since XPS is a surface sensitive technique, it is important to estimate the photoelectron escape depth in the systems under consideration. In accordance with the methodology described in the article [52], the depth at which 95% of the useful signal is formed (information depth) in the [B$_4$C(1.75 nm)/Ru(1.44 nm)]$_{50}$ system with $\Gamma$ = 0.45 is estimated to be approximately 7.7 nm for the B 1s line and 7.4 nm for the Ru 3d line. Since the period of the system is 3.19 nm, it can be assumed that the measured photoelectron spectra carry information from more than two periods. Crucially, spectra contain contributions from both interfaces: Ru-on-B$_4$C and B$_4$C-on-Ru.

Analysis of the decomposed Ru 3d$_{5/2}$ photoelectron spectrum of the multilayer structure [B$_4$C(1.75 nm)/Ru(1.44 nm)]$_{50}$ indicates the appearance of an additional component from the side of lower binding energies relative to pure ruthenium (E = 279.7 eV), which indicates the formation of an additional compound Ru$_x$B$_y$ as a result of the interaction of the Ru and B$_4$C layers. The presence of a shoulder on the side of high binding energies indicates contributions to the spectrum from oxide components. It should be emphasized that due to the small thickness of the main layers, the multilayer structures were not exposed to Ar$^+$ ion sputtering. It is known [53–55] that cleaning with an ion beam can potentially lead to preferential sputtering with significant redistribution of number of atoms within the sample and additional mixing of the layers. Therefore, keeping surface intact in this system is reasonable. The spectrum decomposition indicates the presence of two additional components. The first component is located at a binding energy of 280.5 eV and can be referred to RuO$_2$ [43,56–59] and the second one is centered at the 282.3 eV and can be assigned to RuO$_3$ [43,59–61].

Analysis of the B 1s spectrum of the multilayer [B$_4$C(1.75 nm)/Ru(1.44 nm)]$_{50}$ structure (Fig. 2b) indicates a significant decrease in the intensities of the lines associated with the B-B and B-C bonds and the appearance of a fairly narrow, high-intensity line on the side of lower binding energies in relation to the B-B component, at a binding energy of 187.6 eV. Since in the literature [47,62,63] the peaks of metal borides in B 1s spectra are mostly located at lower binding energies relative to the peak corresponding to the B-B bond, it is reasonable to assume that in the spectrum under consideration the high-intensity peak also reflects the formation of the Ru$_x$B$_y$ compound. The appearance of a peak associated with formation of the borides on the side of lower energies in both the Ru 3d$_{5/2}$ and B 1s spectra is noteworthy. The analysis of the energy position of the peak makes it possible to draw conclusions about the chemical state of the atom in a compound, since the peak position changes depending on the type of surrounding atoms and the nature of chemical bonding. Chemical shifts are primarily caused by the redistribution of the valence electron density when an atom participates in chemical bonding, which directly affects the binding energy of core levels. Different binding energies of electronic levels reflect their charge state in the compound, and significant differences in these values characterize the direction of electron transfer from a less electronegative atom to a more electronegative one. [62,64], in which the shift in binding energy observed in the spectrum is mainly due to a relaxation shift, while the charge transfer in this case is negligible. The relaxation shift can significantly prevail over the chemical shift, resulting in the binding energy shift in the spectrum being mainly determined by the relaxation shift. This preserves the direction of the peak shift in the spectrum, i.e., the peaks reflecting the formation of a compound in the spectra of different components shift in the same direction. It can be assumed that the shift in the binding energy in the Ru 3d$_{5/2}$ spectrum is mainly due to the relaxation shift, and that the charge transfer in this case is insignificant.

Also, the [B$_4$C(2.61 nm)/Ru(0.78 nm)] multilayer structure with $\Gamma = 0.23$ was studied. The calculation of the information depth for this system showed a value of 9.3 nm for the B 1s line and 8.9 nm for the Ru 3d line with a period value of 3.39 nm. Thus, it can be stated that the spectra obtained from this sample also contain contributions from both the Ru-on-B$_4$C interface and the opposite B$_4$C-on-Ru interface. The joint analysis of the Ru 3d$_{5/2}$ and B 1s spectra (Fig. 2) of this system indicates the appearance of an additional low-intensity component in both spectra, which can be attributed to the formation of ruthenium boride as well, but with a different stoichiometry Ru$_m$B$_n$. The additional ruthenium boride Ru$_m$B$_n$ is located at an even lower binding energy than Ru$_x$B$_y$. A stronger shift of the corresponding component towards lower binding energies means a higher electron density on boron, i.e. boron becomes more strongly hybridized with ruthenium. This condition manifests itself in a large number of boron-containing compounds that form complex polyhedral clusters around the metal [65,66].

The significant increase in boron oxide compared to the multilayer structure with the parameter $\Gamma = 0.45$ is noteworthy. It can be assumed that the ruthenium layer of thickness 0.78 nm does not protect the system from oxygen penetration into the structure.

A joint analysis of the spectra obtained from multilayer $B_4C/Ru$ structures with different $\Gamma$ shows that the sample with a thinner Ru layer thickness (smaller $\Gamma$) is prone to the formation of ruthenium borides of different stoichiometry at the interface.

*3.3 System with barrier layer*

One of the perspective ways to limit the formation of transition layers is to apply a thin barrier layer at the interface between the main layers. The use of barrier layers not only reduces the mutual diffusion of the main layers, but also increases the reflectivity by reducing the roughness of the interfaces, as well as improving thermal and radiation stability [26,67]. The material used as a barrier layer must itself have good thermal stability, adhesion, and optical constant values in the operating energy range for a given multilayer mirror.

Carbon was used as the barrier layer material in this work. The use of carbon barrier layers in multilayer X-ray mirrors is a widely adopted practice. It has previously been shown that carbon promotes the smoothing of interfacial boundaries and reduces interdiffusion of the components [68–71]. In this study, the carbon interlayer is applied only at one of the interfaces. In the spectral region near the absorption edge, the empirical Larruquert rule [72,73] is fulfilled, according to which the maximum reflectance of a multilayer structure is achieved for a specific sequence of materials.

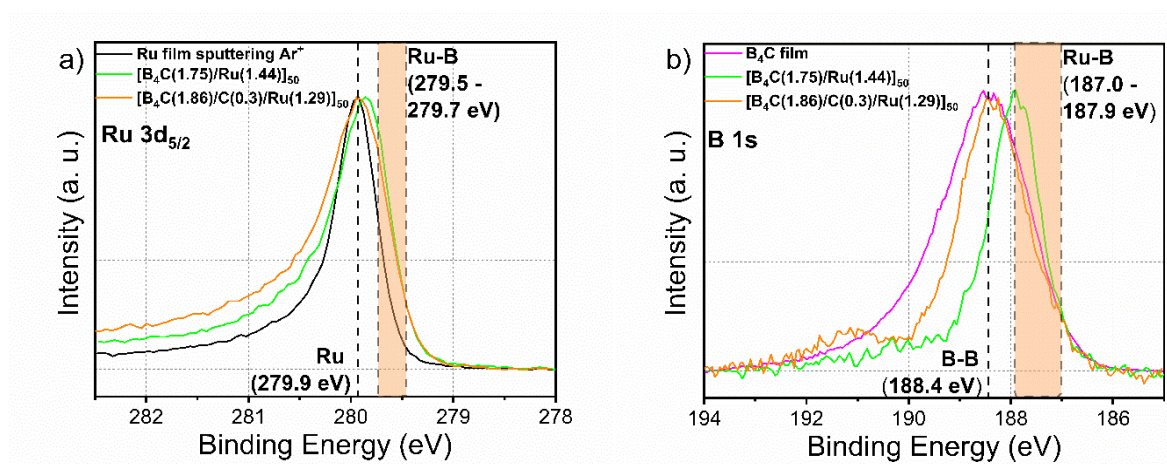

Figure 3. Ru $3d_{5/2}$ (a) and B 1s (b) photoelectron spectra collected from the [B$_4$C(1.75 nm)/Ru(1.44 nm)]$_{50}$, [B$_4$C(1.86 nm)/C(0.3 nm)/Ru(1.29 nm)]$_{50}$ multilayer systems and Ru, B$_4$C reference films.

Figure 3 show the photoelectron spectra of Ru $3d_{5/2}$ and B 1s, respectively, obtained from the [B$_4$C(1.86 nm)/C(0.3 nm)/Ru(1.29 nm)]$_{50}$ and [B$_4$C(1.75 nm)/Ru(1.44 nm)]$_{50}$ multilayer structures and Ru, B$_4$C reference films. The joint analysis of the Ru$3d_{5/2}$ spectra indicates a broadening of the main peak toward higher binding energies in the spectra of the system with the barrier layers compared to the spectrum of the system without barrier layers. In the B 1s spectrum of the system with the barrier layer, the main line exhibits not only broadening but also a noticeable shift. The broadening of the Ru $3d_{5/2}$ peak and the shift accompanied by broadening of the B 1s peak in the system with the barrier layer can be attributed to an increased contribution from pure Ru and B$_4$C, as indicated by the positions of the corresponding reference peaks. One can conclude that the introduction of the C barrier layer into B$_4$C/Ru multilayer structure quite effectively prevents mixing of the main layers and, therefore, significantly reduces the formation of the ruthenium boride compound.

*3.4 Effect of high-temperature annealing*

As noted in the introduction, the main application field of B$_4$C/Ru multilayer mirrors is optical systems for EUV lithographers. Since some mirrors in the optical systems are exposed to X-ray and extreme ultraviolet radiation of high intensity (up to 250 W [6]) for a long time, they can be subjected to severe thermal load. Heating of optical elements can lead to undesirable disruption of periodicity of multilayer structures due to mixing of layers, crystallization, etc. and, as a consequence, reduce their reflectivity. Therefore, mirrors in EUV lithographs must be heat resistant, i.e., withstand long-term exposure to high temperatures without substantial changing their original characteristics. Additionally, annealing the multilayer structure at high temperatures enhances the intermixing of the main layers, which allows one to study the diffusion properties of atoms in a multilayer coating.

Despite the fact that some studies [22,23] report minimal changes in the reflectivity of multilayer B$_4$C/Ru mirrors during annealing up to 550 °C, other authors [74] report that already at a temperature of 490 °C, a sharp degradation of the multilayer structure occurs, which is accompanied by a strong decrease in reflectivity. Thus, the question of the thermal stability of the B$_4$C/Ru structure remains open. The authors of the work [74] claim that at a temperature of 400 °C, the thickness of the transition junctions at the interfaces begins to grow. However, a catastrophic loss of reflective properties has not yet occurred. Thus, 400 °C is the "threshold" temperature for studying the initial stages of degradation of the B$_4$C/Ru structure. Based on these data, the [B$_4$C(1.75)/Ru(1.44)]$_{50}$ system with Γ = 0.45 and d = 3.19 nm was annealed at 400 °C for 1 hour to study thermal stability at this temperature.

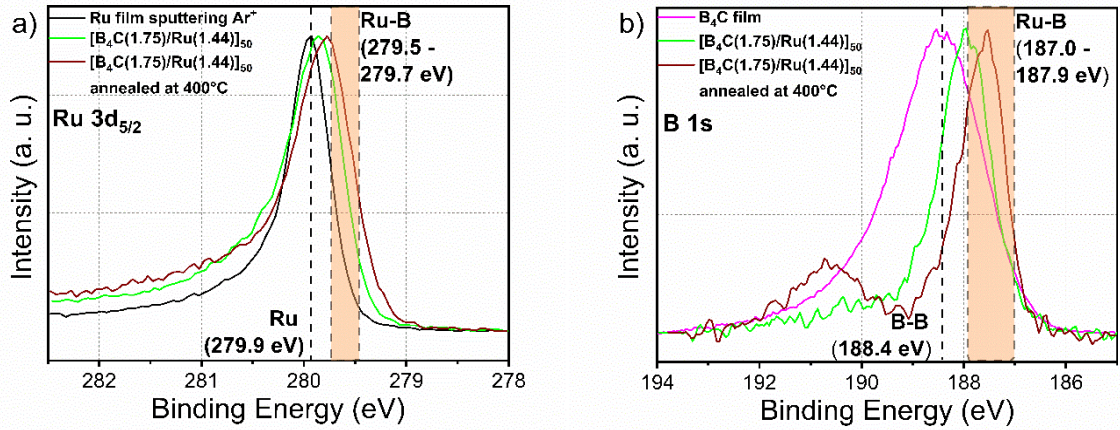

Figure 4. Ru 3d$_{5/2}$ (a) and B 1s (b) photoelectron spectra collected from the Ru and B$_4$C reference films and [B$_4$C(1.75 nm)/Ru(1.44 nm)]$_{50}$ multilayer system before and after annealing at 400 °C.

Figure 4a shows the Ru 3d$_{5/2}$ photoelectron spectra measured from [B$_4$C(1.75 nm)/Ru(1.44 nm)]$_{50}$ multilayer system before and after annealing, as well as the spectrum of the reference Ru film. It can be seen that annealing at 400 °C affects the shape and position of the main peak. In particular, there is a noticeable shift and broadening of the main peak towards lower binding energies. This fact indicates active mixing of the main layers during thermal annealing and, as a result, more active formation of Ru-B bond. Similar trend (without broadening) is observed in the B 1s spectra presented in Figure 4b, also indicating increased intermixing of the main layers in the annealed multilayer structure. A strong shift of the B 1s peak towards lower binding energies by 0.4 eV is traced in the B 1s spectrum of the annealed multilayer structure confirming the formation of great amount of the ruthenium boride. Also, a noticeable growth of the boron oxide peak is observed in the B 1s spectrum, which indicates active oxygen movement deep into the structure under the Ru surface layer.

It can be concluded that annealing at 400 °C for an hour enhances the mixing of layers, increasing the content of the formed ruthenium boride. The multilayer structure Mo/B$_4$C, which is also often considered as a system for use in lithography, has a significantly higher thermal stability [75,76].

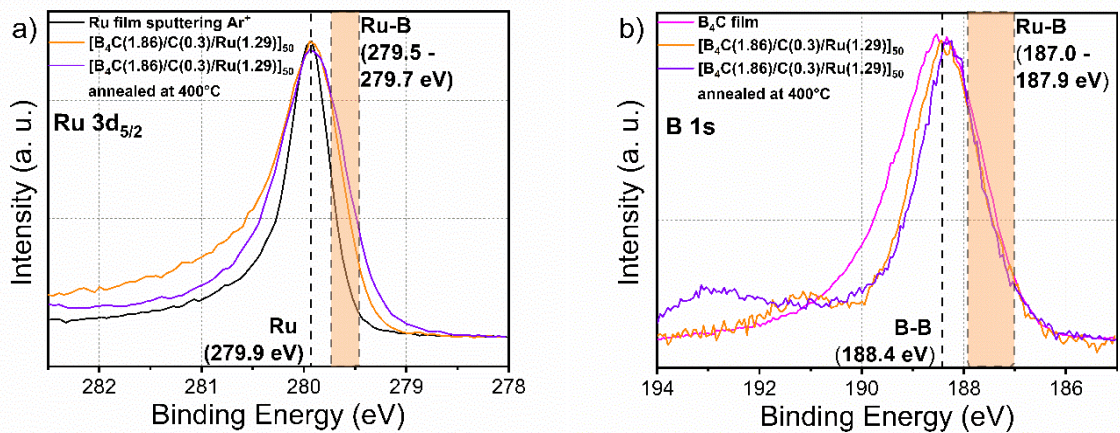

Figure 5. Ru 3d$_{5/2}$ (a) and B 1s (b) photoelectron spectra collected from Ru and B$_4$C reference films and [B$_4$C(1.86 nm)/C(0.3 nm)/Ru(1.29 nm)]$_{50}$ multilayer system before and after annealing at 400 °C.

The introduction of a thin C barrier layer 0.3 nm thick on the Ru-on-B$_4$C interface completely changes the situation. The [B$_4$C(1.86 nm)/C(0.3 nm)/Ru(1.29 nm)]$_{50}$ sample with was also annealed at a temperature of 400 °C for 1 hour. Figure 5a shows Ru 3d$_{5/2}$ photoelectron spectra obtained from pure reference Ru, as well as the [B$_4$C(1.86 nm)/C(0.3 nm)/Ru(1.29 nm)]$_{50}$ system before and after annealing. Despite the formation of a slightly higher intensity shoulder at the main peak from the lower binding energies after annealing, there is no noticeable shift of the peak towards lower binding energies. Thus, it can be seen that in the B$_4$C/Ru system with the C barrier layer, in contrast to the system without a barrier layer, annealing at a temperature of 400 °C does not lead to the increased formation of ruthenium boride. The analysis of B 1s photoelectron spectra (Figure 5b) confirms the conclusions reached. Unlike a sample without a barrier layer, in which, after annealing, there was a strong shift of the main peak towards lower binding energies, there is no noticeable shift of the main peak in the system with the barrier layer.

It is worth noting the behavior of ruthenium and boron oxides components in the spectrum of this system. During annealing, the intensity of the shoulder of the main Ru 3d$_{5/2}$ peak from high binding energies, which indicates the presence of RuO$_2$ and RuO$_3$ oxides, decreases. At the same time, a 1.5 eV shift of the oxide peak towards higher binding energies is observed in the B 1s spectrum. This effect can be explained by the fact that during annealing, oxygen previously bound to ruthenium migrates to more underlying layers and binds to boron. Thus, a boron oxide with higher oxygen content is formed in the structure after annealing, which is manifested in a higher binding energy of this peak.

Thus, it can be concluded that the system with the barrier layer is more thermally stable than a multilayer structure without the barrier layer. If annealing at 400 °C in the system without the barrier layer leads to its noticeable degradation, then in the system with a barrier layer, similar annealing practically does not affect the state of the interfaces.

*3.5 XRD and XRR results*

Diffraction patterns were obtained for all the structures studied to determine the presence/absence of crystallization in the samples (Figure 6). For clarity, the diffraction curves have been shifted vertically. Crystallization can negatively affect the reflectivity of multilayer mirrors. Crystalline materials, as a rule, exhibit anisotropy of growth, which leads to heterogeneity of the layers. On the other hand, amorphous structures tend to have smoother and sharper layer boundaries. Thus, it is important to preserve amorphous structures. All the peaks observed in Figure 6 relate exclusively to the silicon substrate. Thus, it can be concluded that all structures retain an X-ray amorphous state.

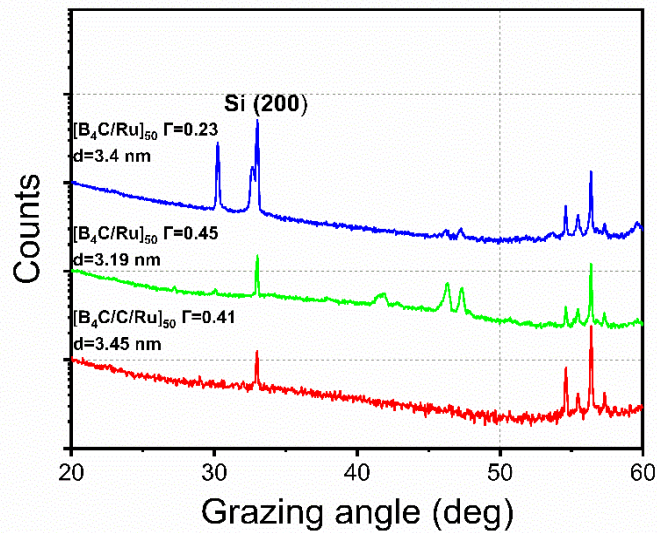

Figure 6. Diffraction patterns measured for the studied $B_4C/Ru$ (without and with C barrier layer) multilayer structures.

To determine the thickness of the main and transition layers, in all studied multilayer structures the measurements using X-ray reflectometry (XRR) were performed. Figure 6 shows the experimental (black dots) and fitted (red lines) reflection curves. For better visual perception, the reflection curves were shifted vertically. The captions of curves in the figure indicate the nominal values of the periods (d) and parameters of Γ.

A theoretical model was constructed based on the X-ray photoelectron spectroscopy data discussed above to fit the experimental curves. The models took into account surface layers containing oxygen (in the form of Ru and B oxides) and carbon contaminants, which required the subsurface region to be considered separately from the underlying layers, represented as a periodic structure. In the multilayer system, the variation in layer thickness (Δz) with increasing period number, caused by the heterogeneity of deposition during growth, was taken into account. To describe the interlayer transition regions, a parameter $s$ was introduced, combining the contributions of geometric roughness (including growth roughness) and mixing zones with the formation of the ruthenium boride. Various physical processes in the mixed regions were modelled by different functional dependencies: interdiffusion between the main layers was approximated by an error function, while mixing caused by energetic particle bombardment during sputtering was described by a linear dependence. Geometric roughness was described by an exponential function [77]. During the optimisation of the model parameters, the densities of the materials were fixed in accordance with the reference values from the Springer Materials database and were not varied [78]. Figure 7 shows the results of adjustments of the theoretical curves to the experimentally obtained results for B$_4$C/Ru samples with different parameter Γ, as well as with the introduced barrier layer C. Table 4 shows the approximation parameters obtained for the samples presented.

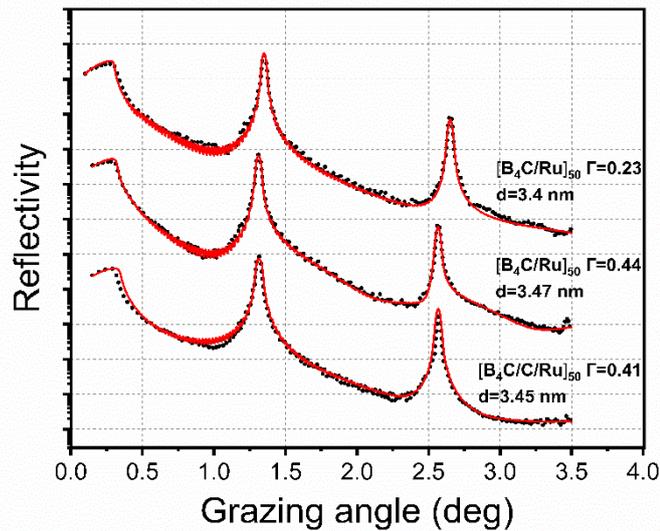

Figure 7. Experimental reflectance curves (black dots) and corresponding approximations (red lines) for the multilayer structures [B$_4$C/Ru].

Table 4. Fitting parameters for the models of the multilayer samples (ρ – density, z - layer thickness, Δz – thickness drift, d – period thickness, Γ - ratio of $z_{Ru}$ to d, s – transition region thickness (includes roughness and diffuseness)).

| Layer | N | ρ, g/sm³ | z, nm | Δz, % | s, nm |
|---|---|---|---|---|---|
| [B$_4$C/Ru]$_{50}$ d = 3.4 nm, Γ = 0.23 | | | | | |
| RuO$_2$ | 1 | 6.97 | 0.3 | | 0.06 |
| Ru | 1 | 12.3 | 0.48 | | 0.05 |
| B$_2$O$_3$ | 1 | 1.85 | 0.6 | | 0.6 |
| B$_4$C | 1 | 2.51 | 2.01 | | 1.2 |
| Ru | 49 | 12.3 | 0.92 | 0.53 | 0.4 |
| B$_4$C | | 2.51 | 2.43 | -0.22 | 0.5 |
| [B$_4$C/Ru]$_{50}$ d = 3.47 nm, Γ = 0.44 | | | | | |
| RuO$_2$ | 1 | 6.97 | 0.83 | | 0.7 |
| Ru | 1 | 12.3 | 0.69 | | 0.53 |
| B$_2$O$_3$ | 1 | 1.85 | 1.5 | | 0.12 |
| B$_4$C | 1 | 2.51 | 0.44 | | 0.39 |
| Ru | 49 | 12.3 | 1.37 | 0.98 | 0.36 |
| B$_4$C | | 2.51 | 2.1 | -0.65 | 0.47 |
| [B$_4$C/C/Ru]$_{50}$ d = 3.45 nm, Γ = 0.41 | | | | | |
| RuO$_2$ | 1 | 6.97 | 0.1 | | 0.1 |
| Ru | 1 | 12.3 | 1.31 | | 0.38 |
| C | 1 | 1.7 | 0.31 | | 0.05 |
| B$_2$O$_3$ | 1 | 1.85 | 0.3 | | 0.05 |
| B$_4$C | 1 | 2.51 | 1.44 | | 1.2 |
| Ru | 49 | 12.3 | 1.53 | 0.29 | 0.12 |
| C | | 1.7 | 0.3 | 0.01 | 0.25 |
| B$_4$C | | 2.51 | 1.64 | -0.27 | 0.25 |

As can be seen from Table 4, the values of parameter s for the B$_4$C and Ru layers show a significant difference in the transition layer thicknesses between the near-surface period and the main multilayer structure in all samples studied. This discrepancy may be due to the ambiguity of the solutions to the inverse reflectometry problem, which means that the fitting parameters must be interpreted with caution. Nevertheless, a clear systematic dependence is observed: for two samples with different values of Γ, the parameter s in the multilayer part of structures remains practically constant, amounting to ~ 0.47 – 0.5 nm for B$_4$C and ~ 0.36 – 0.4 nm for Ru. This suggests that the variation in the Γ parameter does not have a significant effect on the thickness of the mixed layer and roughness in the B$_4$C/Ru system. A comparison of the sample with Γ = 0.44 and the sample with a carbon barrier layer (Γ = 0.41) revealed a decrease in the s parameter by 0.22 – 0.24 nm for both materials after the introduction of the barrier layer. The results obtained indicate a positive effect of the carbon barrier layer, leading to a decrease in both the thickness of the interface regions and the roughness value.

*4. Conclusion*

A comprehensive study of multilayer $B_4C/Ru$ structures with different parameters Γ have been carried out using X-ray photoelectron spectroscopy (XPS), X-ray diffraction (XRD) and X-ray reflectometry (XRR).

Active interaction of the main layers in the system with the formation of ruthenium borides has been established, and an increase in the thickness of the $B_4C$ layer (a decrease in the Γ parameter) leads to the formation of ruthenium borides of different stoichiometry. A ruthenium layer of thickness 0.78 nm does not protect the $B_4C$ layer from oxidation.

The introduction of an ultrathin (0.3 nm) carbon barrier layer at the Ru-on-$B_4C$ interface leads to a significant redistribution of the peak intensities of the components responsible for metallic Ru (in the Ru $3d_{5/2}$ spectrum) and B-B and B-C bonds (in the B 1s spectrum), which indicates an inhibition of the formation of ruthenium boride.

The thermal stability of the $B_4C/Ru$ system was studied during annealing at 400 °C for 1 hour before and after the introduction of the C barrier layer. The results of the XPS studies of the multilayer structure without a barrier layer showed a significant increase in the mutual diffusion of the layers, accompanied by an increase in the thickness of the mixing regions and oxidation of the surface layers. This indicates a limited thermal stability of the $B_4C/Ru$ structure. The introduction of a C barrier layer on the Ru-on-$B_4C$ interface increases the thermal stability of the system. Annealing at 400 °C leads only to a redistribution of oxygen in the system. Thus, the use of multilayer $B_4C/Ru$ systems with a barrier layer is preferable in optical systems exposed to prolonged radiation exposure.

The analysis of the XRD curves shows the X-ray amorphousness of all the studied samples. The analysis of reflection curves by the XRR method reveal a decrease in the parameter s (thickness of the transition region) by 0.22–0.24 nm in the system with barrier layer.

The results obtained are important for the development of highly efficient multilayer mirrors used in EUV lithography and X-ray optics. This work demonstrates the prospects of using $B_4C/Ru$ systems with a carbon barrier layer to minimize the effects of the interaction of the main layers and the formation of sharp interfaces in these structures to create high-quality optical elements.

**Funding sources**

Russian Science Foundation Grant No. 19-72-20125-П.

**Acknowledgements**

This work was supported by the Russian Science Foundation (RSF), grant No. 19-72-20125-П. The authors thank the staff of the "Centre for X-ray Diffraction Studies" and the "Centre for Physical Methods of Surface Investigation" of the Research Park at Saint Petersburg State University for their assistance with XPS, XRD, and XRR measurements. We also gratefully acknowledge the National Research Center "Kurchatov Institute" for providing access to the ESCA laboratory facilities and for their valuable technical support, which greatly contributed to this research.